\documentclass[final]{article}

\usepackage{booktabs} 

\usepackage{nips_2018}

\usepackage[utf8]{inputenc} 
\usepackage[T1]{fontenc}    
\usepackage{hyperref}       
\usepackage{url}            
\usepackage{booktabs}       
\usepackage{amsfonts}       
\usepackage{nicefrac}       
\usepackage{microtype}      

\usepackage{graphicx}
\usepackage{amsmath}
\usepackage{caption}
\usepackage{subcaption} 
\usepackage{csquotes}
\usepackage{bm} 
\usepackage{appendix}
\usepackage{hyperref}

\newtheorem{definition}{Definition}

\begin{document}
\title{A simulation based dynamic evaluation framework for system-wide algorithmic fairness}

\author{
  Efr\'en Cruz Cort\'es \\ 
  University of Colorado School of Public Health\\
  \texttt{efren.cruzcortes@gmail.com}
\And
 Debashis Ghosh \\
 University of Colorado School of Public Health \\
 \texttt{debashis.ghosh@ucdenver.edu}
}

\maketitle

\begin{abstract}
We propose the use of Agent Based Models (ABMs) inside a reinforcement learning framework in order to better understand the relationship between automated decision making tools, fairness-inspired statistical constraints, and the social phenomena giving rise to discrimination towards sensitive groups. There have been many instances of discrimination occurring due to the applications of algorithmic tools by public and private institutions. Until recently, these practices have mostly gone unchecked. Given the large-scale transformation these new technologies elicit, a joint effort of social sciences and machine learning researchers is necessary. Much of the research has been done on determining statistical properties of such algorithms and the data they are trained on. We aim to complement that approach by studying the social dynamics in which these algorithms are implemented.  We show how bias can be accumulated and reinforced through automated decision making, and the possibility of finding a fairness inducing policy. We focus on the case of recidivism risk assessment by considering simplified models of arrest. We find that if we limit our attention to what is observed and manipulated by these algorithmic tools, we may determine some blatantly unfair practices as fair, illustrating the advantage of analyzing the otherwise elusive property with a system-wide model. We expect the introduction of agent based simulation techniques will strengthen collaboration with social scientists, arriving at a better understanding of the social systems affected by technology and to hopefully lead to concrete policy proposals that can be presented to policymakers for a true systemic transformation.
\end{abstract}


\section{Introduction}
Advances in machine learning have proven to be useful for disciplines such as medicine, public health, climate change, poverty research, and many other disciplines in the natural and social sciences. Many governmental institutions and private companies have bought into the promise that machine learning can solve all problems with enough data. As such, there is a trend towards automation of important decision-making procedures that affect the lives of individuals and communities. The desire for automation may be driven out of pure profit interests, as is the case with insurance and real estate companies.  Alternatively, governmental institutions and the criminal justice system wish to improve societal conditions through use of these types of algorithms. However, implementation of such algorithms has proven to be detrimental in many cases, a major cause being the reproduction and reinforcement of undesirable patterns in the data. Data which are already biased due to historic discrimination and other structural deficiencies lead to `biased' machine learning algorithms. See \cite{oneill2016weapons, eubanks2018automating} for further discussion of this point.  

A notable and now-classic example of particular interest to health policy makers is the unchecked implementation of machine learning algorithms in the criminal justice system. The 2016 report by ProPublica \cite{angwin2016machine} studied one such algorithm used in Broward County, Florida, which intended to predict recidivism among defendants up for parole. The report found out the algorithm was reproducing systematic bias against people of color and in favor of whites. The creators of the algorithm, however, argued their algorithm was fair since it satisfied equal positive predictive values (see \cite{chouldechova2017fair}).  In the public eye, reports like this unveiled the erroneous assumption that algorithmic tools are neutral.  This has led to the growth of interest in studying algorithmic fairness. For the research community, one major area of focus has been on defining measures of fairness and designing algorithms to satisfy such measures. Since then, many notions of fairness have been quantified into statistical properties, see for example, \cite{berk2018fairness} and \cite{narayanan2018twentyone}. In general, not all statistical quantifications of fairness are compatible (e.g. \cite{chouldechova2017fair}), and for now the best practice is to choose the most suitable one to the problem at hand knowing possible drawbacks in other directions.

In this paper, as a working example, we develop a very simple model of an arrest-recidivism system. We model the dynamics of agents we call ``cops'' and of two demographic groups in a population. We will see that even if the algorithmic tool itself (in this case a simple classifier) satisfies certain notions of fairness, the system may fail to do so. The use of ABMs help us realize that phenomenon as well as to find the reason behind it.

We briefly discuss related work in Section \ref{sec:related_work}, provide the rationale and structure for a system-wide analysis framework in Section~\ref{sec:system_wide}, introduce a simple example model in Section \ref{sec:model_intro} and analyze it in Section~\ref{sec:model_analysis}. We proceed to explore the model in terms of causality in Section~\ref{sec:causal}, and use reinforcement learning to find optimal policies in Section~\ref{sec:reinforcement}.
We conclude with some future research directions in Section \ref{sec:conclusion}.

\subsection{Related Work}\label{sec:related_work}

Our approach will consider both recidivism prediction and predictive policing. The report \cite{angwin2016machine,larson2016analyzed} showed the widely-used COMPAS algorithm violates certain statistical notions of fairness, and \cite{chouldechova2017fair}, showed an inherent mathematical trade-off among certain notions of fairness when considering heterogenous groups. In \cite{dressel2018accuracy} COMPAS is shown to be as accurate and fair as a simple linear model and an aggregate of people with little criminal justice expertise, rebutting the belief that these prediction tools, if imperfect, are fairer and more accurate than decisions taken by humans. The work in \cite{lum2016predict} shows that predictive policing algorithms reinforce historical police activity, resulting in a suboptimal process through which the algorithm is incapable of accurately estimating crime rates. A survey of crime forecasting procedures is given in \cite{berk2013statistical}, while a survey on statistical notions of fairness for algorithmic tools in the criminal system is given in \cite{berk2018fairness}. For a civil rights perspective on the problem, see \cite{goodman2018algorithms}. The recent book \cite{ferguson2017rise} provides a comprehensive overview of tools, interventions, effects, and context of algorithmic use in policing and surveillance. Finally, \cite{friedler2018comparative} provides a framework to compare different proposed fairness metrics.

\section{A System-Wide Analysis Framework}\label{sec:system_wide}

While fairness and justice have served as catalysts for the new wave of sound research regarding algorithm accountability, the fundamental structures and dynamics in which these same fairness and justice concepts are reified get lost in the background. In the majority of previous studies, authors have focused on statistical properties of the algorithms being implemented, acknowledging the inherent problem in the data and data collection strategies but not always consolidating such processes as part of the system of analysis. This concern is particularly relevant for policy makers, since they ideally will rely on our expertise to take influential decisions. As researchers, we need to be wary of finding optimal tools for structural problems and not just to improve already established tools.  There might be fundamental flaws in the working assumptions of these tools that lead to undesirable consequences. While the concept of ``algorithm neutrality'' vanishes, the concept of ``algorithm fairness'' arises, which more often than not conveys a similar message, namely that it is possible for fairness to be achieved by an algorithm.  A ``fair algorithm'' can be well defined as an algorithm satisfying some quantification of a fairness notion. ``Fairness on a population'', however, which is determined by the consequences of a procedure of which the algorithm is part of, is not a property of the algorithm, but of the population itself, and can only be satisfied and understood if the algorithm and the consequences of its use are placed in context. To cite Victor Hugo:

\begin{displayquote}
The guillotine is the concretion of the law; it is called vindicte; it is not neutral, and it does not permit you to remain neutral.... All social problems erect their interrogation point around this chopping-knife. ... [T]he scaffold is not a machine.... It seems as though it were a being, possessed of I know not what sombre initiative; ... possessed of will.
\end{displayquote}
\cite{hugo1887miserables}. Using this analogy is a crude yet enlightening example. Notwithstanding the efforts to make algorithms fair, many of these algorithms are themselves harmful weapons and we may be sharpening their blade. For example, recidivism prediction tools are counterfactually punitive, working on a double illusion: that a particular individual's characteristics are those of the statistical aggregate of their inputted population, and that punishment based on predictions of human actions are in any way philosophically founded. This might be an obvious example, but more furtive algorithms are out there supporting dubious practices, as it is the case, for example, with targeted advertisement of junk food to children, where the problem is not how fair it is but that it exists at all. The same goes for financial and economic research, where relations of exploitation are not questioned by algorithm makers, focusing only on profit. Indeed, such is the hold of utilitarianism and economic reason on our research that we still use concepts like ``utility function'' to refer to human goals. This paper does not claim in any way to successfully overcome utilitarian logic, or to avoid the pitfall of sharpening blades, but we hope awareness and conversation can lead us in the right direction. We make an effort to contextualize algorithms in order to understand the effects of automation on a social system, keeping in mind, again, Hugo's words:
\begin{displayquote}
The scaffold is the accomplice of the executioner; it devours, it eats flesh, it drinks blood; the scaffold is a sort of monster fabricated by the judge and the carpenter, a spectre which seems to live with a horrible vitality composed of all the death which it has inflicted.
\end{displayquote}

\subsection{System wide framework}\label{sec:system_awareness}

Our intention in this paper is not to introduce a new notion of fairness but a new framework for analysis. We intend to do so by introducing the use of Agent Based Models (ABMs) to the community, and aim to demonstrate their usefulness in discovering system-wide problems and solutions, as well as elusive causal mechanisms. Three principles guide this paper: systemic analysis, causal relations and mechanisms, and optimal interventions. {\bf Systemic analysis} refers to setting the algorithm into a context. This implies analyzing the data generating process, the decision making stage, and its consequences all under the same framework. Most previous research does set its analysis into context and warn of its consequences, however we strive to analyze these portions of the system together with the algorithm itself, subject to the same quantification, processing, and statistical measures. This principle provides a structure to the system and an engine for its overall dynamics. {\bf Causal relations and mechanisms} are imperative in our understanding of how (un)fairness arises in a particular system. Note the difference between causal relations and causal mechanisms. A causal relation indicates what variable has a causal effect on another variable, a causal mechanism explains how this effect is produced. Causal relations are discovered under causal inference frameworks in statistics; in this paper we use ABMs as resources for causal mechanism discovery. Finally, although intervention effects are part of causal analysis, it is not always straightforward to find {\bf optimal interventions} when the system under consideration is too complex or computationally demanding, so we give them their own special stage. In this paper we take advantage of the nature of ABM simulations as counterfactuals and embed our ABM in a reinforcement learning framework in order to find optimal policies, or parameter combinations, through a principled process.

\begin{figure}[t]
\centering
\includegraphics[width=.5\textwidth]{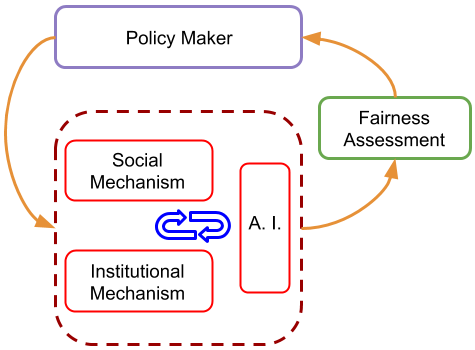}
\caption{System Loop}
\label{fig:system_loop}
\end{figure}

We propose a system-wide framework based on a reinforcement learning paradigm. Under this scenario, the agent will be the policy maker and the environment a conglomerate of social, institutional, and technological mechanisms, as shown in Figure~\ref{fig:system_loop}. In the model, the three mechanisms inside the dashed stage interact in any general way, and may do so for a while before an output is observed. Notice we have explicitly marked in the schematic a ``fairness assessment'' stage, as different definitions of fairness may yield different equilibrium points of the system. Indeed, most previous research on ``fair machine learning'' regards this stage of the process.

Take as an example the social phenomenon of ``street crime'' and the current institutional actions around it. Two particular mechanisms have been of interest in the literature: police surveillance and bail grants. In the schematic the social mechanism would be all social aspects which relate to crime, the institutional mechanism pertains to police surveillance and arrests, as well as the eventual judicial process, and the AI mechanism would be any tool involved in relevant decision making, for example predictive policing and recidivism assessment algorithms. Notice this particular case is that of ``street'' crime and not crime in general, since other forms of potentially more harmful crime, like war, corporate, and environmental crimes, are policed in different ways, if policed at all.

Previous work has studied some of the relations inside the environment stage, in particular regarding the AI and institutional mechanisms. The work in \cite{ensign2017runaway} treats the system of predictive policing as a feedback loop modeled by a P\'olya urn. They further develop regret bounds for such feedback scenarios in \cite{ensign2018decision}. Notice in our schematic that this inner loop is different from the outer loop, which allows for policy interventions by the global policy maker agent.

Currently, machine learning systems are used to optimize and speed up decision processes previously made by humans. As such, one may wonder if the structure of the system is fundamentally affected at all by AI tools. Indeed, technologies and infrastructures are intertwined with and often etiological to social relations. For example, historical transformations of family and gender relations are affected and catalyzed by material conditions (\cite{maynes2012family}, \cite{federici2004caliban}). Indeed, not going too far into the past, everyone knows how the internet has transformed modern economies and conceptions of the self. As eloquently put by a well-known economist (pardon the use of ``man'' as a universal):
\begin{displayquote}
Technology reveals the active relation of man to nature, the direct process of the production of his life, and thereby it also lays bare the process of the production of the social relations of his life, and of the mental conceptions that flow from those relations.
\end{displayquote}
\cite{marx1990capital}. Hence, we believe a thorough examination of social and artificial mechanisms is necessary for a true transformation into fair and ethical societies.

\subsection{ABM rationale}\label{sec:abms}
Collecting data from and intervening on social systems is an expensive and time-consuming process; in many situations it is impossible due to ethical or logistic constraints.  Computational models provide a feasible avenue to understand social systems provided they exhibit similar characteristics. ABMs are a successful simulation paradigm in that they can generate a plethora of social dynamics and allow for policy experimentation.

ABMs differ from other models, for example differential equation-based models, in that they admit more flexible heterogeneity in their populations (see \cite{rahmandad2008heterogeneity}). Each agent has its own characteristics and often acts according to its local environment and imperfect information. This behavior is more realistic than many classical models of human behavior where perfect knowledge and rationality were assumed, as well as a homogeneity in the population.

We have placed a lot of emphasis on the ability to identify causal relations. ABMs are particularly useful to this task in that they provide simple generative mechanisms to an observed phenomenon and each simulation run can be interpreted as a counterfactual (\cite{macy2009social}, \cite{marshall2014formalizing}). A general motto for agent-based modeling is ``simpler is better'', as the distillation of complex dynamics into smaller parts provides better insight into the system. In a sense, these explanatory mechanisms can be thought of as the simplest dynamics necessary to generate our observation. Occam-like induction philosophies assert these simple mechanisms are the most probable actual causes, see for example \cite{solomonoff1964formal}.

A general overview of the usefulness and history of agent-based models in the social sciences can be found in \cite{macy2002factors}. The book \cite{epstein2006generative} contains numerous examples of ABMs in the social sciences, their design, and study. Finally, \cite{wilensky2015introduction} is a step by step introduction to the design of ABMs through the popular software Netlogo.

\section{A Simple Example Model}\label{sec:model_intro}
In a general sense, we can view ABMs as a set of agents following simple rules of interaction, usually with imperfect or local knowledge. Although the interaction are simple and local, complex global patterns can emerge, providing insights regarding causal hypotheses  of a particular system's observed behavior.

\begin{figure}[t!]
\centering
\begin{subfigure}{.4\textwidth}
\centering
\includegraphics[width=\textwidth]{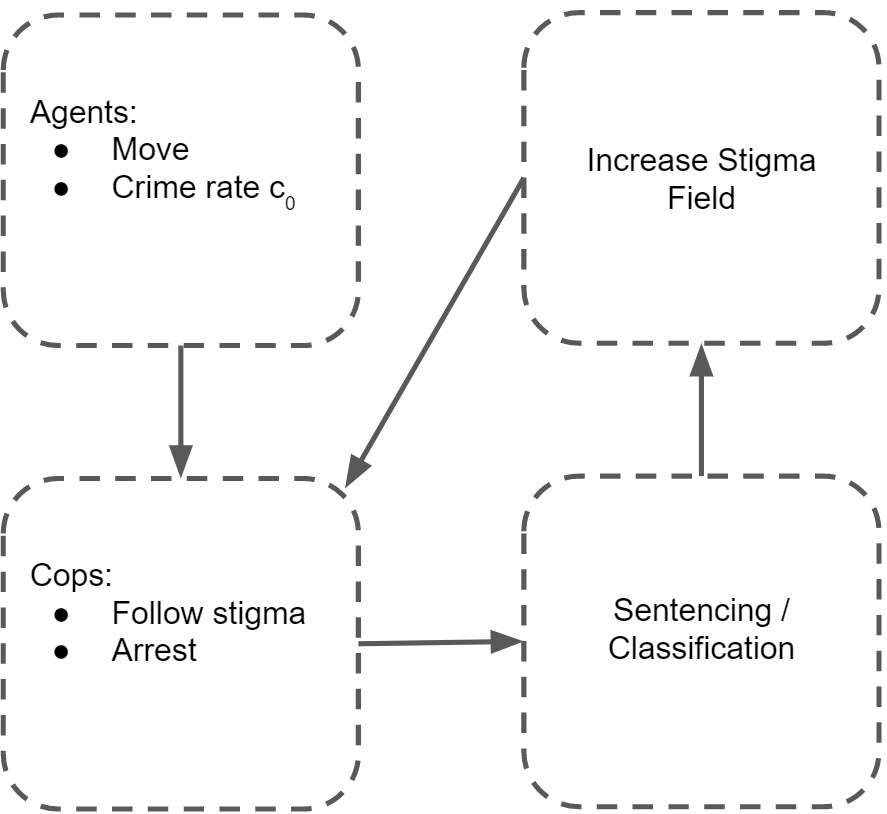}
\caption{Simple schematic of the order of events in the agent based model.}
\label{fig:model_diagram}
\end{subfigure}
~
\begin{subfigure}{.4\textwidth}
\centering
\includegraphics[width=\textwidth]{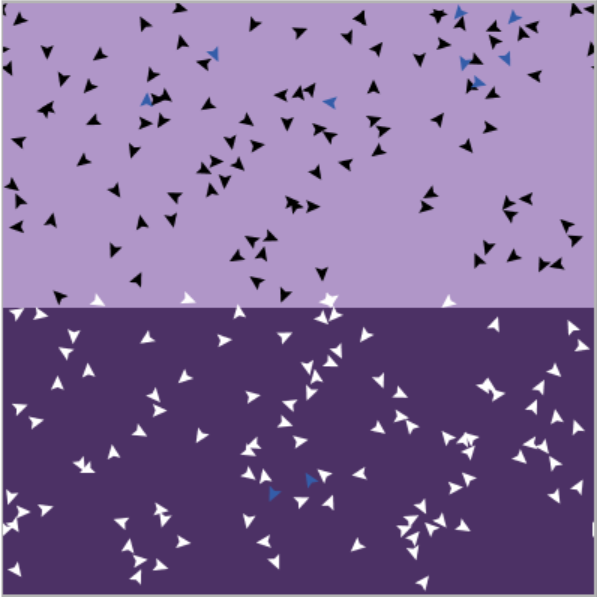}
\caption{Sample original configuration of the model. Members different groups are confined to different regions. Cops travel freely among regions but $q_0 \%$ of them start in the first region.}
\label{fig:original_configuration}
\end{subfigure}
\caption{Agent Based Model}\label{fig:abm_schematics}
\end{figure}

To understand the arrest-sentence system of interest we start as simple as possible. We model each member of the population as an independent agent that randomly moves in a confined region with demographically similar agents. Besides their membership to one of two groups, their crime and recidivism rates, agents do not have other characteristics. We also have a few cops moving, but these cops interact with population agents when a crime is witnessed, producing an arrest. This model is simple and certainly does not capture all nuances in the true dynamics of a population, however, it serves as a foundation to understand population-level phenomena that may arise from simple assumptions and interactions. A schematic of the system's dynamics is shown in Figure~\ref{fig:model_diagram}.

The model is composed of a grid of cells, {\bf the world}, in which {\bf agents} move and act. The agents are divided into cops and two populations groups, $G_1$ and $G_2$. The two populations commit crime at a constant rate $c_0$, and if a cop is present the crime-committing individual from the population will get arrested. Cops, who are unevenly distributed among the populations at the beginning of the simulation, move ``following their noses'' according to a {\bf stigma field} which places bias in locations with crime history and reinforces repeated surveillance. The probability with which a cop will follow the stigma field is denoted $\theta$ and it will have an important role in the subsequent analysis. Once a population agent is arrested they undergo a trial stage, for which the trial decision (i.e., go to jail or not) is based on a random classifier. Notice we have set the crime and recidivism rates constant across groups (see Appendix for explanation of why). We will see that even under such assumption an original placement bias of police can lead to large disparities. For details on the model, its dynamics, and, most importantly, the justification for assumptions made, please see the Appendix. Code is found in \url{https://github.com/efrencc/F-abm}.

Figure \ref{fig:original_configuration} shows the original configuration of agents for a sample run. In this case the model was run using Netlogo software.

\section{Model Analysis}\label{sec:model_analysis}

We will use two basic notions of fairness as described by \cite{chouldechova2017fair}: Predictive Parity and Error Rate Balance. We define them slightly different than in \cite{chouldechova2017fair}, as they were defined using score outcomes and we use a hard $0/1$ classifier outcome. These conditions are based on the false positive and false negative rates (FPR, FNR), as well as the positive predictive value (PPV) of the score or classifier, which only witnesses people who have been arrested. Our analysis considers the entire system and not just the sentencing stage so that we make explicit the condition ``arrest''. In the following $A$ stands for the arrest variable, $R$ for true recidivism, $J$ is the output of the classifier, and $G$ is the group membership.

\begin{definition}[A-conditioned Predictive Parity] A system implementing classifier $J$ satisfies {\it A-conditioned predictive parity} if the probability of true recidivism, given a positive classifier assignment and given $A$, is the same across groups. That is:
\begin{align*}
\mathbb{P}(R = 1 | J = 1, A = 1, G = G_1) = \mathbb{P}(R=1 | J = 1, A = 1, G = G_2) .
\end{align*}
We refer to these probabilities as $PPV_{A}(g)$.
\end{definition}

\begin{definition}[A-conditioned Error Rate Balance] A system implementing classifier $J$ satisfies {\it A-conditioned error rate balance} if the False Positive and False Negative Rates, given $A$, are the same across groups. That is:
\begin{align*}
\mathbb{P}(J = 1 | R = 0, A = 1, G = G_1) = \mathbb{P}(J = 1 | R = 0, A = 1, G = G_2)
\end{align*}
for the FPR, and
\begin{align*}
\mathbb{P}(J=0 |R=1, A=1, G = G_1) = \mathbb{P}(J=0 |R=1, A=1, G = G_2)
\end{align*}
for the FNR. We refer to these probabilities as $FPR_{A}(g)$ and $FNR_{A}(g)$.
\end{definition}

Notice two main differences from \cite{chouldechova2017fair}: First, we talk about a system implementing a classifier, as opposed to the classifier itself. Second we condition on the arrest variable $A$. We consider to appropriately study fairness in a sentencing system we need to study not just the algorithmic tool but also the data fed into it, this is, the dynamics that generated such data.

Equation (2.6) of \cite{chouldechova2017fair} describes the relationship among PPV, FPR, FNR, and prevalence $p_A(g) = \mathbb{P}(R=1 |G=g,A=1)$, and we reproduce it here for the sake of completeness:
\begin{align}\label{eq:chouldechova}
FPR_A(g) = \frac{p_A(g)}{1-p_A(g)}\frac{1 - PPV_A(g)}{PPV_A(g)}(1-FNR_A(g)).
\end{align}

As stated in \cite{chouldechova2017fair}, if the prevalence differs across groups, we cannot obtain ERB and PP simultaneously. If prevalence is equal between the groups, however, it is possible to satisfy all of these fairness metrics.

Let's now explore our system in light of these equations. It is easy to show that for our model $FPR_A(g) = r_c$, $FNR_A(g) = 1 - r_c$, and $PPV_A(g) = r_0$ (see Appendix). Hence we achieve both A-conditioned ERB and A-conditioned PP.

While this outcome is satisfactory for the classifier itself (in terms of fairness metrics, not accuracy), it doesn't provide information about system-wide fairness.

To assess overall fairness in our model we can add the deviations from the ideal case, by defining the quantity
\begin{align*}
\tau_A := \left|1 - \frac{PPV_A(G_1)}{PPV_A(G_2)}\right|
 + \left|1 - \frac{FPR_A(G_1)}{FPR_A(G_2)}\right|
 + \left|1 - \frac{FNR_A(G_1)}{FNR_A(G_2)}\right|.
\end{align*}
In the ideal case the numerator and denominator of each component are equal and therefore $\tau_A=0$. We could have looked at the absolute difference instead of the ratio but there are two future advantages of this definition. In the first place, we will soon define analogous population quantities which involve some terms that are hard to estimate from the model. When we consider ratios, however, those particular terms cancel out. Secondly, in Section~\ref{sec:causal}, we will study tolerance values of these $\tau$ quantities, in this instance it will be more intuitive and interpretable to define these fairness metrics in terms of ratios.

\begin{figure}[t!]
\centering
\begin{subfigure}[t]{.4\textwidth}
\includegraphics[width=\textwidth]{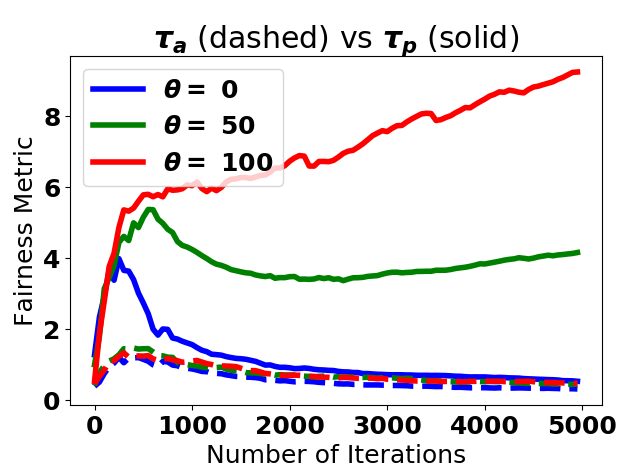}
\caption{Arrested and population measures $\tau_A$ (dashed) and $\tau_P$ (solid). As $\theta$ increases the system becomes less fair, but it is only evidenced through $\tau_p$.}
\label{fig:tau_total}
\end{subfigure}\hspace{4mm}%
~
\begin{subfigure}[t]{.4\textwidth}
\centering
\includegraphics[width=\textwidth]{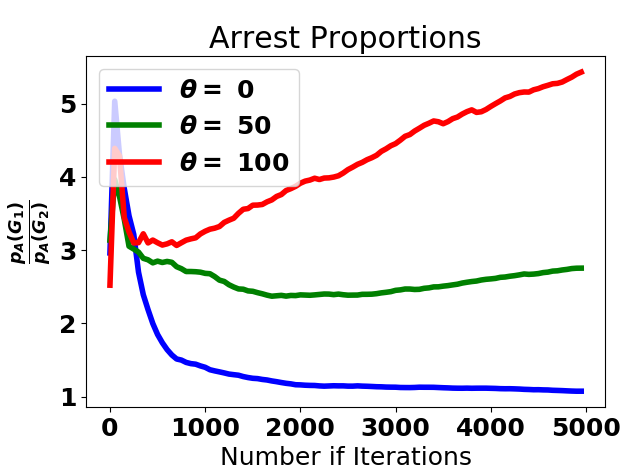}
\caption{Proportion of arrested members of group $G_1$ to arrested members of group $G_2$. Notice that for high values of $\theta$ this proportion can be extremely large.}
\label{fig:arrest_proportion_simple}
\end{subfigure}
\caption{Fairness assessment comparison and arrest proportion, the missing link.}
\end{figure}

In Figure \ref{fig:tau_total} we plot the averaged value of $\tau_A$ (dashed lines) over 30 simulations for different values of the parameter $\theta$. $\tau_A$ approaches zero as the system stabilizes. However, as we will shortly see, there is still an unfair process not captured in this: the proportion of arrests among different groups.

In Figure \ref{fig:arrest_proportion_simple} we plot the ratio of arrested members of $G_1$ to arrested members of $G_2$, averaged over 30 runs. This time we note the arrest rates are highly disproportionate. In particular low values of $\theta$ (the probability with which cops follow the prior stigma of a neighborhood) lead to almost equal arrest rate. Large values of $\theta$, however, can lead to ever-increasing extreme disparities between both arrest rates.

Disproportionate arrest rates are not justified when crime rates are the same and should be considered part of the fairness assessment. The observed difference could be explained by differences in crime rates; however, as already mentioned, the agents in the model commit crime at constant rate independent of group membership.  There is another element driving these ratios upward. Recognizing this element motivates us to introduce the population fairness metrics:

\begin{definition}[Population Error Rate Balance] A system implementing classifier $J$ satisfies {\it population error rate balance} if the False Positive and False Negative Rates, defined population-wise, are the same across groups. That is:
\begin{align*}
\mathbb{P}(J = 1 | R = 0, G = G_1) = \mathbb{P}(J = 1 | R = 0, G = G_2)
\end{align*}
for the FPR, and
\begin{align*}
\mathbb{P}(J=0 | R=1, G = G_1) = \mathbb{P}(J=0 | R=1, G = G_2)
\end{align*}
for the FNR. We refer to these probabilities as $FPR_{P}(g)$ and $FNR_{P}(g)$.
\end{definition}

\begin{definition}[Population Predictive Parity] A system implementing classifier $J$ satisfies {\it population predictive parity} if the probability of true recidivism, population wise and given a positive classifier assignment, is the same across groups. That is:
\begin{align*}
\mathbb{P}(R = 1 | J = 1, G = G_1) = \mathbb{P}(R=1 | J = 1, G = G_2) .
\end{align*}
We refer to these probabilities as $PPV_{P}(g)$.
\end{definition}

Notice Population PPV equals A-conditioned PPV, as it is easy to show $PPV_P(g) = \mathbb{P}(R = 1 | J=1, G=g, A=1)$ and hence population and arrest PP are equal conditions in this case (see Appendix for details). We now define a new overall measure of fairness, this time for the population:
\begin{align*}
\tau_P := \left|1 - \frac{PPV_P(G_1)}{PPV_P(G_2)}\right|
 +\left|1 - \frac{FPR_P(G_1)}{FPR_P(G_2)}\right|
 + \left|1 - \frac{FNR_P(G_1)}{FNR_P(G_2)}\right|.
\end{align*}

Figure \ref{fig:tau_total} shows $\tau_P$ averaged over 30 runs for several values of $\theta$. This time we can easily see the discrepancy of results. If we consider how the implementation of the algorithms affects the whole population, taking into account the context under which data is collected, bias surfaces to light.

By adapting equation \ref{eq:chouldechova} to the new case, we obtain:
\begin{align*}
FPR_P(g) = \frac{p_P(g)}{1-p_P(g)}\frac{1 - PPV_P(g)}{PPV_P(g)}(1-FNR_P(g)),
\end{align*}
where the population prevalence $p_P(g) = \mathbb{P}(R=1 |G=g)$.

Again, ERB cannot be achieved because the prevalence is different, indeed, the prevalence is
\begin{align*}
p_P(g) &= \mathbb{P}(R=1 |G=g,A=1)\mathbb{P}(A=1 | G=g)
+  \mathbb{P}(R=1 |G=g,A=0)\mathbb{P}(A=0 | G=g) \\
& = \mathbb{P}(R=1 |G=g,A=1)\mathbb{P}(A=1 | G=g),
\end{align*}
which reveals a new culprit $\mathbb{P}(A=1|G=g)$, the probability of arrest. Since this probability varies by group $G$, prevalence also depends on $G$. Remember, however, that cops in our model did not discriminate when they observed crime, and both groups had the same crime rate. What then leads to the different arrest probability? Indeed,
\begin{align*}
\mathbb{P}(A=1 | G) & = \mathbb{P}(A=1 |C=1, G)\mathbb{P}(C=1|G)
+ \mathbb{P}(A=1|C=0, G)\mathbb{P}(C=0|G) \\
& = \mathbb{P}(A=1|C=1, G)\mathbb{P}(C=1|G),
\end{align*}
where we have assumed the cops do not arrest when there is no crime or at disproportionate rates, although, unfortunately, there is well-documented evidence of disparities in arrest rates among different demographic groups as well as arrests and convictions of innocent people (see for example \cite{innocence_project}, \cite{sentencing_project} and the many articles in \cite{nytimes}). We do not, however, treat such case here. There are two reasons for which arrest probability would be different, crime rate and arrest probability given a crime is committed. We can similarly further break up $\mathbb{P}(A=1|C=1, G)$ conditioning on the cop being present, and it is then that the surveillance rate is revealed.

We now conclude that even if anything else in the system is ``fair'', dissimilar surveillance rates propitiate unfair outcomes. As our model shows, there is no need for cops to discriminate themselves, nor to surveil differently across groups. The only thing in our model enforcing systemic discrimination is that cops follow the historic stigma of a region, and that in the initial conditions there is a higher distribution of cops in the first group's region.

\section{Causal Analysis}\label{sec:causal}
As mentioned in the introduction, a guiding principle for the paper is to provide possible causal mechanisms in addition to causal relations. Since the structure of ABMs is mechanistic in nature, ABMs are good candidates to fulfill this principle. For them to be considered causal mechanisms, however, depends on the causality framework under consideration. The affinity to ascribe causal mechanisms for ABMs is set by any causal framework formulated around counterfactual outcomes (see \cite{hedstrom2010causal}). Indeed, each simulation run (given a specified set of initial conditions) can be considered a counterfactual (see \cite{marshall2014formalizing}), and it is under this framework that we explore the outcomes of our model.

\begin{figure}[t!]
\centering
\begin{subfigure}{.4\textwidth}
	\centering
	\includegraphics[width=1\textwidth]{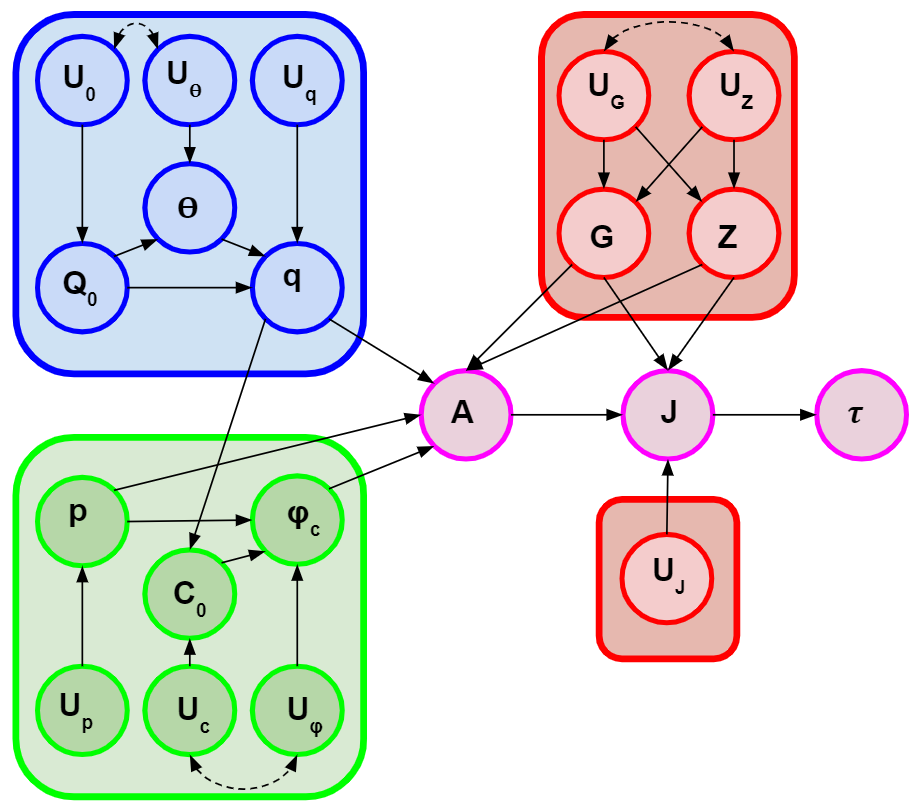}
	\caption{Full causal graph}
	\label{fig:causal_graph}
\end{subfigure}\hspace{4mm}%
~
\begin{subfigure}{.4\textwidth}
	\centering
	\includegraphics[width=1\textwidth]{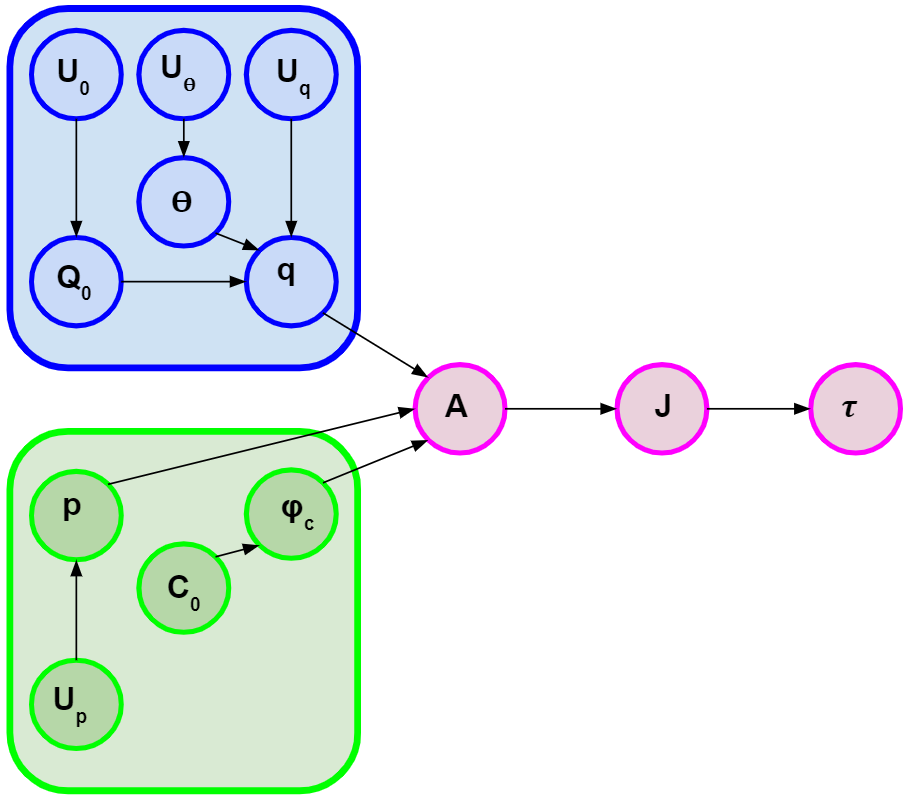}
	\caption{Causal graph for ABM model}
	\label{fig:causal_graph_model}
\end{subfigure}
\caption{Full and ABM causal graphs.}\label{fig:big_causal_graphs}
\end{figure}

A full causal graph would at least include the relationships shown in figure~\ref{fig:causal_graph}. The $\mathbf{U}$ variables represent endogenous variables, $(\mathbf{p}, \mathbf{q})$ are the coordinates of civilians and cops, respectively, $\mathbf{c_0}$ is a variable crime rate while $\mathbf{\psi_c}$ is a binary crime indicator, $\mathbf{Z}$ are personal characteristics and $Q_0$ is the bias with which sops are place on the first zone at the beginning. We have, of course, made some assumptions in our ABM model that simplify the full graph. For example, the crime rate is held constant, while arrests by cops depend only on having committed a crime, and the probability with which cops will surveill with stigma is independent of their current location. The graph pertaining to the model is shown in figure~\ref{fig:causal_graph_model}.

\begin{figure}[t!]
\centering
\begin{subfigure}{.25\textwidth}
	\centering
	\includegraphics[width=\textwidth]{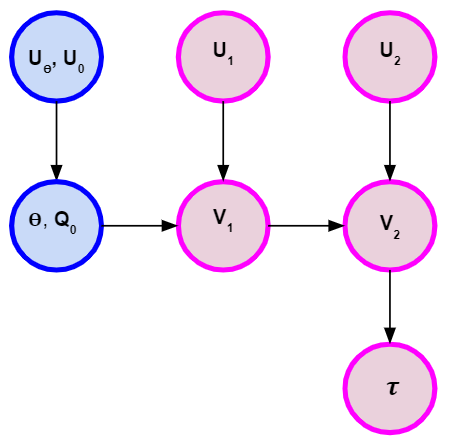}
	\caption{Simplified causal graph}
	\label{fig:causal_graph_simplified}
\end{subfigure}\hspace{4mm}%
~
\begin{subfigure}{.25\textwidth}
	\centering
	\includegraphics[width=\textwidth]{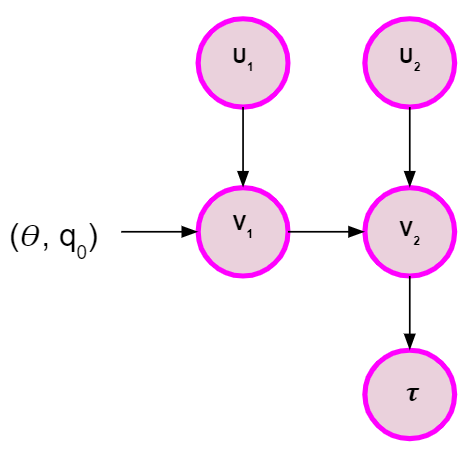}
	\caption{Simplified causal graph with intervention}
	\label{fig:causal_graph_simplified_do}
\end{subfigure}
\caption{Simplified graph and intervention model.}\label{fig:simple_graphs}
\end{figure}

Since our outcome of interest is the variable $\tau$, after intervening on $Q_0$ and $\bm{\theta}$, we can group all other variables into variables $\mathbf{U}$ and $\mathbf{V}$ and hence further simplify the graph as shown in figure~\ref{fig:causal_graph_simplified}. We are now ready to consider different intervention outcomes. We do this by setting particular values of $\bm{\theta}$ and $Q_0$. The resulting model is shown in figure~\ref{fig:causal_graph_simplified_do}, which provides us with the distribution of interest:
\begin{align}
\mathbb{P}(\tau | do(\bm{\theta} = \theta), do(Q_0 = q_0) ),
\end{align}
see \cite{peters2017elements}, and \cite{pearl2009causal}. Therefore, we can estimate causal effects from the outcome distributions.

\begin{figure}[t]
\centering
\includegraphics[width=.95\textwidth]{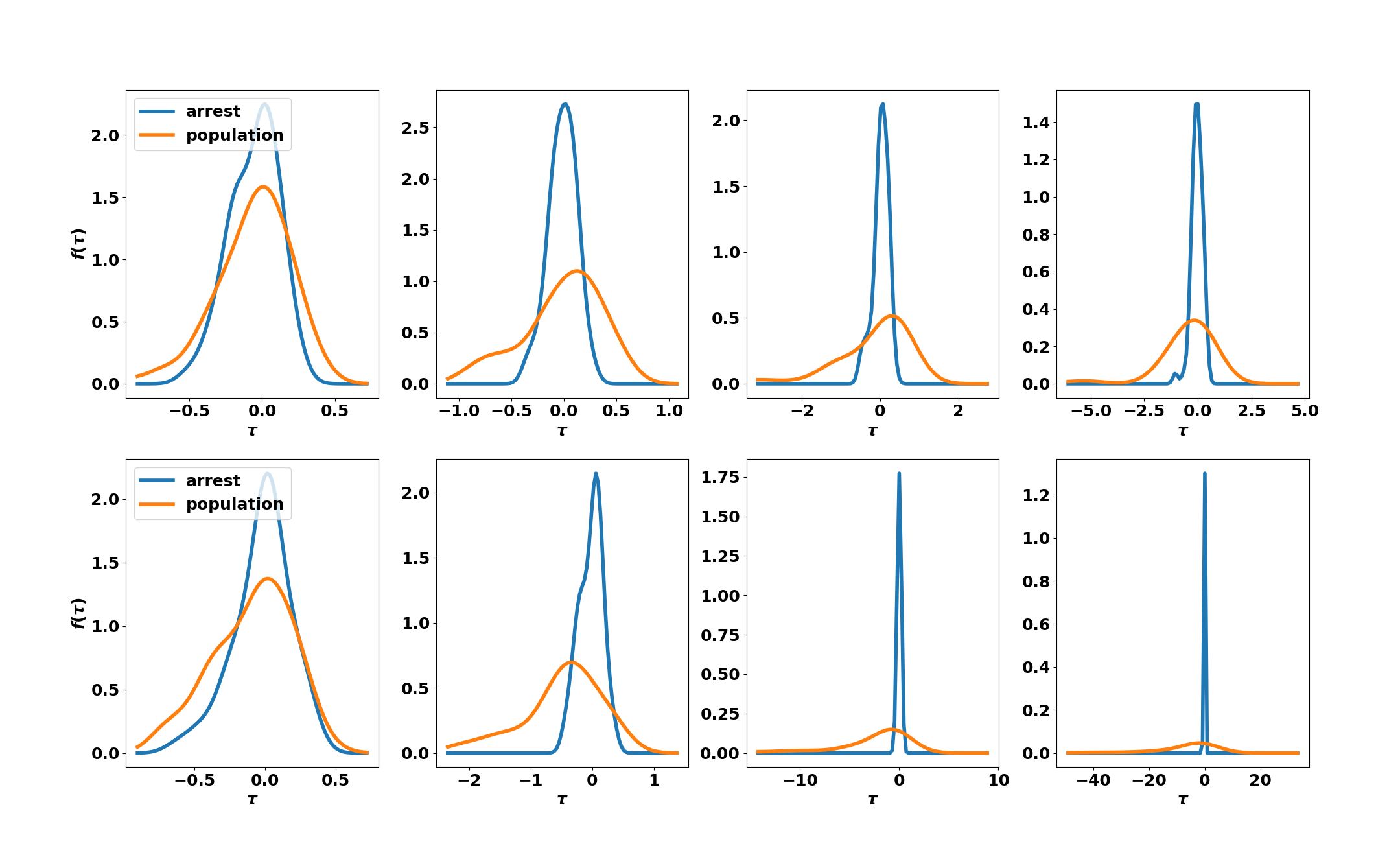}
\caption{distributions of outcomes for $q_0=.5$ (first row) and $q_0 = .8$ (second row). Each column is a value of theta, from low to high, in \{0, .25, .75, 1\}. }
\label{fig:outcome_distributions}
\end{figure}

For simplicity we focus on $\tau^1_A : = 1 - \frac{FPR_A(G_1)}{FPR_A(G_2)}$, and $\tau^1_P$, defined similarly. To estimate these outcomes for different values of $\theta$ and $q_0$ we ran a simulation of the model for $5000$ steps and computed the outcomes at the last step. We then averaged over $60$ of these runs. We chose $q_0\in\{.5,.8\}$ and $\theta\in\{0, .25, .5, .75, 1\}$. Figure~\ref{fig:outcome_distributions} shows the kernel density estimates of the outcomes for the different combinations of $(q_0,\theta)$. The first row shows the results for $q_0=.5$ and the second those of $q_0 = .8$.

\begin{figure}[t]
\centering
\begin{subfigure}{.4\textwidth}
	\centering
	\includegraphics[width=\textwidth]{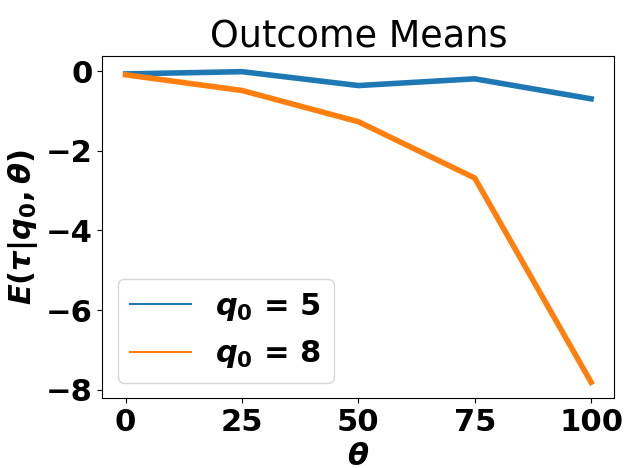}
	\caption{Outcome means for population}
	\label{fig:outcome_means_population}
\end{subfigure}
~
\begin{subfigure}{.4\textwidth}
	\centering
	\includegraphics[width=\textwidth]{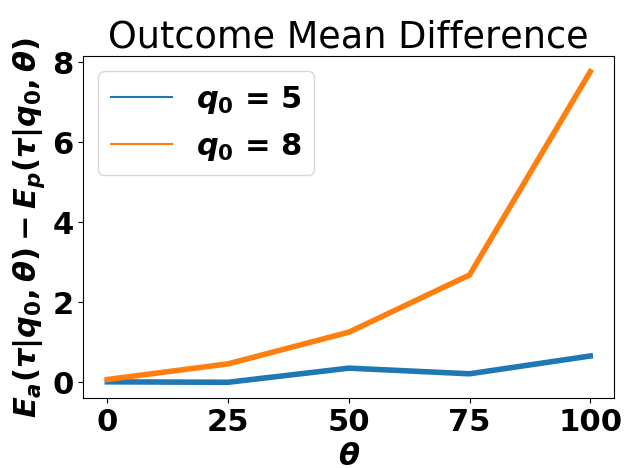}
	\caption{Difference between population and arrested outcome means}
	\label{fig:outcome_mean_difference}
\end{subfigure}
\caption{Comparison between outcome means for arrested and overall populations.}\label{fig:outcome_means}
\end{figure}

We notice that independent of the value of $(q_0,\theta)$, $\tau^1_A$ concentrates much more around its means than $\tau^1_P$. Furthermore, this mean is close to zero (desired outcome), as expected. This is not surprising in light of the results of section~\ref{sec:model_analysis}. What is surprising is that such stark difference prevails even for the ``fair'' initial allocation of cops dictated by $q_0 = .5$ (first row). The means of $\tau^1_P$ given $q_0=.5$ are indeed close to zero (see figure~\ref{fig:outcome_means_population}), however, there is a nonnegligible amount of mass away from zero. This is because with an even distribution of cops on the two group neighborhoods, converging into a state of high stigma differences is equally likely across groups, but it remains a likely situation. That is, the mere dynamics of stigmatic surveillance make unfair population outcomes likely.

The case $q_0 = .8$ is more drastic. We again have a great amount of mass away from zero, but also the means have shifted away from zero, showing a negative bias towards the non-privileged group. Figure~\ref{fig:outcome_means_population} clarifies this. In the figure we can see how rapidly the means move away from zero whenever $\theta > 0$. A comparison of the outcomes means between $\tau^1_P$ and $\tau^1_A$ is shown in figure~\ref{fig:outcome_mean_difference}.

\begin{figure}[t]
\centering
\includegraphics[width=.75\textwidth]{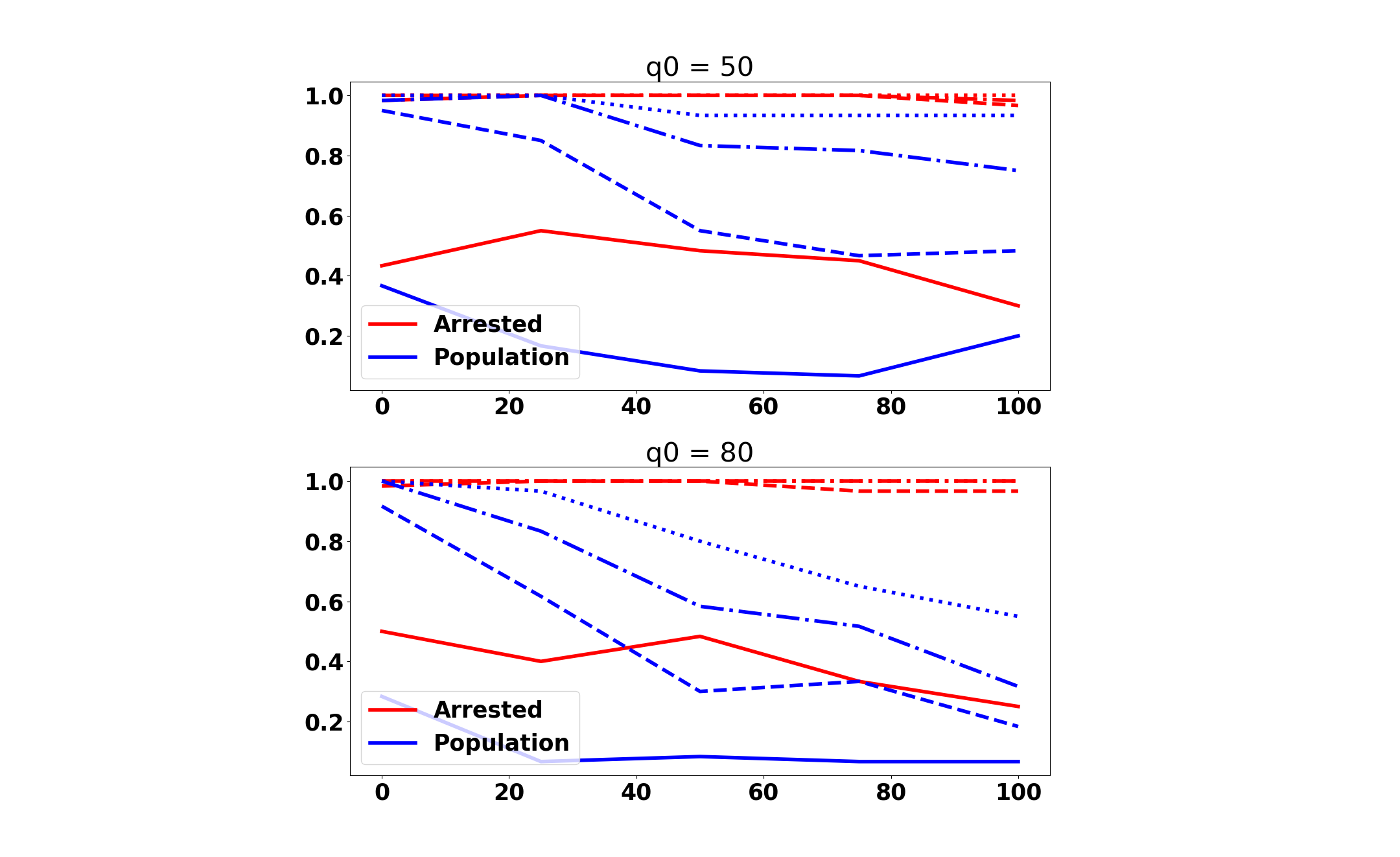}
\caption{Proportion of outcomes considered to be ``fair'' according to tolerances in \{.1, .5, 1, 2 \}. We compare the arrested (red) and population (blue) cases for (a) $q_0 = .5$ and (b) $q_0 = .8$.}
\label{fig:fairness_proportions}
\end{figure}

Practically, the raw values of $\tau^1$ might not be as useful as a conceptualization of their meaning as indicators of ``fairness'' in the system. For this purpose we consider a tolerance threshold $\epsilon_{tol}$ and look at the variable $Y = \chi_{\{|\tau^1| < \epsilon_{tol}\}}$. This variable will indicate if the system is within $\epsilon_{tol}$ of ``perfect fairness'', regardless how much $\tau^1$ deviates from zero. If $\epsilon_{tol} = 1$, for example, it indicates $FPR(G1)$ is at most twice as big as $FPR(G2)$. Figure~\ref{fig:fairness_proportions} shows the proportion of outcomes considered ``fair'' under tolerance values $\epsilon_{tol} \in \{.1, .5, 1, 2 \}$. For ease of comparison, we plot them in different ways.

In the first quadrant we see that unless $\epsilon_{tol}$ is too small, for the arrested case we achieve ``fairness'' most of the time, independent of the initial cop distribution $q_0$. This is not surprising. On the second quadrant we see the same comparison for the population-wide case. As $\epsilon_{tol}$ grows the $q_0=.5$ case achieves large proportions of fair outcomes even for large $\theta$. For $q_0=.8$, however, there is a large proportion of unfair outcomes even for large $\epsilon_{tol}$s. The third and fourth quadrants are redundant of the first two, and show that independent of $q_0$ the arrested outcome will always seem to be fair unless $\epsilon_{tol}$ is very small, hiding the unfair cases revealed by the population outcomes.

\section{Reinforcement Learning}\label{sec:reinforcement}
The last guiding principle revolves around optimal interventions. To begin understanding how we could satisfy such principle we setup a simple multi-armed bandit problem. For a background on this problem see \cite{sutton2018reinforcement}. We focus on the case $q_0 = .5$, $\epsilon_{tol}$, and $\tau^1_P$. The set of actions to be taken is the set of possible values for $\theta$: $\{0, .25, .5, .75, 1\}$. The rewards are the outcomes $Y$ from section~\ref{sec:causal}. For fixed $(\theta, q_0,\epsilon_{tol})$, the distribution of $Y$s is dictated by the distributions in figure~\ref{fig:outcome_distributions} as shown in figure~\ref{fig:fairness_proportions}.

\begin{figure}[t]
\centering
\begin{subfigure}{.4\textwidth}
	\centering
	\includegraphics[width=\textwidth]{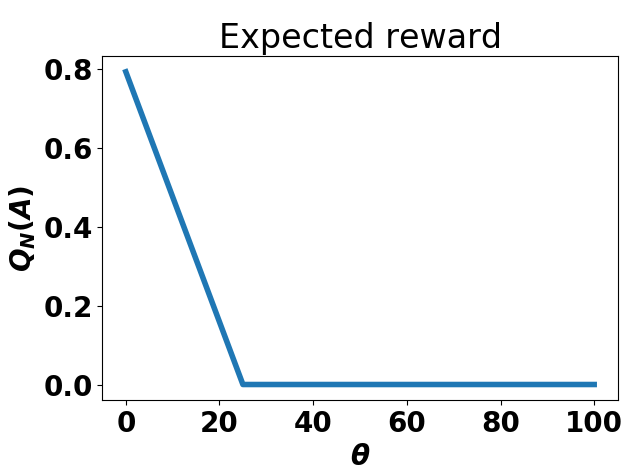}
	\caption{Expected reward over $30$ runs of the simulation.}
	\label{fig:expected_reward}
\end{subfigure}
~
\begin{subfigure}{.4\textwidth}
	\centering
	\includegraphics[width=\textwidth]{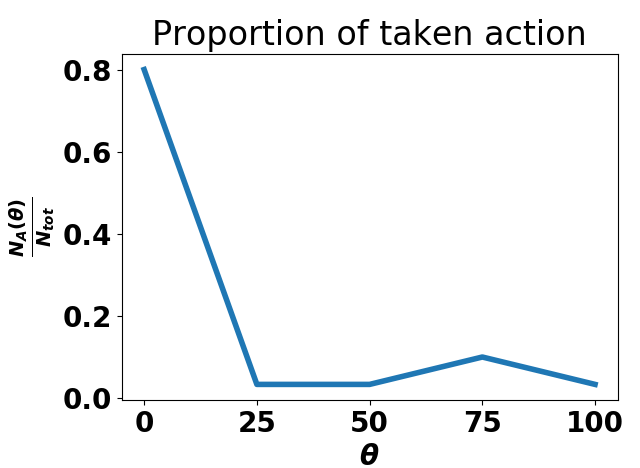}
	\caption{Proportion of an action taken by $\epsilon$-greedy algorithm.}
	\label{fig:action_proportions}
\end{subfigure}
\caption{Expected reward and proportion of actions taken according to these rewards.}\label{fig:rl_stats}
\end{figure}

The goal is to explore values of $\theta$s and design a policy to choose the $\theta$ that maximizes expected reward. We used an $\epsilon$-greedy algorithm with $\epsilon = .1$ and with value function of an action equal to the average reward obtained when that action was chosen. Figure~\ref{fig:expected_reward} shows the expected reward for different actions over $30$ runs of $3000$ steps each. Figure~\ref{fig:action_proportions} shows the proportion of times a specific action was taken. It does not take many runs for the algorithm to realize the optimal action is $\theta = 0$, that is, completely unbiased surveillance. Similar results hold for different values of $q_0$.

This simple example shows a reinforcement learning scenario is able to produce an optimal policy choice when dealing with this particular ABM model. Although simple, the purpose of this example is to be a precursor for more complicated settings. In reality, ABMs will have more than one parameter, or ``action'', for which we'll want to create a policy. For example, we could introduce a ``social worker'' agent, a policy would then involve both $\theta$ as well as the distribution and behavior of these new social workers, even a trade-off between the two if there are limited resources and only a number of them can be put to work. Similarly, we only considered independent runs starting at time zero, but in general we'd like to know how outcomes change when we change parameters without resetting the model. That is, what if cops have been surveilling under $\theta_1$ for a while, and then they change to $\theta_2$? How do outcome distributions change? In these cases we'll have to consider both possible stationarity and  multiple environment states. The more we refine our basic AB model the more the need to have an automated way to learn an optimal policy. We believe reinforcement learning is a good candidate for such an enterprise. 

\section{Conclusion and Future Work}\label{sec:conclusion}

We have presented a simple agent-based model with the aim of understanding some elements that give rise to inequality in an arrest-sentence system. With the aid of the model we discovered discrimination can occur at the population level, even if this discrimination is not apparent when studying the algorithmic tool in isolation. We defined population level versions of different fairness metrics and noticed that when we look at those metrics the model indeed shows highly disparate results. An important takeaway is that under very simple assumptions about agent and cop behavior, including constant crime and recidivism rate, as well as unbiased arrest in the presence of crime, discrimination can still arise as a consequence of prior history. We also discovered that the best way be de-bias the system is for the cops to ignore the stigma of neighborhoods, and follow a random path.

The model was made simple on purpose, with the objective to show that even simple models can give rise to enlightening phenomena. Future directions of research include adding a few elements to the system that make it more realistic. For example, endowing the population agents with simulated covariates that resemble real data. By doing so, we open the possibility to implement other classifiers and understand their effect in the overall system.


\appendix
\section*{Appendix}
\subsection*{Simple arrest model and its assumptions}\label{sec:model_details}
\subsubsection*{The World.}
The world, which is a grid of cells in which agents can move and act, is divided in two regions, accounting for the neighborhoods in which our two types of agents (besides cops) interact. Cops will be able to move freely between regions, while the two population types are assumed restricted to one region, with no interaction among them. The assumption of such separation is a strong one, however, in the case of race, such constraint reflects the way most neighborhoods have statistically skewed demographics. Indeed, an early and notable example of ABM's aimed to explain racial segregation \cite{schelling1969models, schelling1971dynamic}. There are no further constraints in the original setup of the world.

\subsubsection*{The population agents.}
In our model there are two types of population agents: members of group $G_1$, and members of group $G_2$, where the groups are taken to represent a distinction over one or more sensitive or protected variables. For example, group $G_1$ could be a minority as defined by socially arbitrary racial categorizations, while group $G_2$ would be the majority or privileged group (say, white defendants). Besides position $\mathbf{p}^{(i)}_t$ of agent $i$ at time $t$, their parameters are:

{\it Crime rate}. The rate $c_0$ at which agents commit crime. In this toy model such rate is kept constant across individuals, independent of group membership. Note that although in many cases true crime rates are not known, there is reason to believe some actions considered crime are constant across demographics groups, for example, drug use \cite{lum2016predict}.

{\it Recidivism rate}. The rate $r_0$ at which individuals recidivate. Although this parameter is hard to estimate from real data, numerous studies provide evidence that prison sentences do not affect likelihood of rearrest, [*** See Nagin and Snodgrass 2013, \cite{ridgeway2019experiments}], allowing us to make the simplified assumption that the classifier's decision and the iteration time won't affect $r_0$.

\subsubsection*{Cop agents.}
Cops are allowed to move freely among group neighborhoods. They are initially unevenly distributed among the populations, with a ratio of $q_0$ in the discriminated group. We encode the cops positions in the vector $\mathbf{q}$. Cops also have the following parameters:

{\it Arrest rate given crime observed}. In this simple model we also simplify the behavior of cops by assuming a constant arrest rate given crime observed $r_a$. Notice that this simplification is in general unrealistic as racial bias is well-documented among police (\cite{lum2016predict}, \cite{alexander2012new}). However, there are instances in which race, at least directly, is not an influencing factor for arrest, as is the case of traffic violation and the so called ``veil of darkness'', \cite{ridgeway2019experiments}. Note, however, that a backdoor path is possible through, for example, car make, year, and well-kept status. In the simplest of cases we have set $r_a = 1$.

{\it Surveillance bias}. Cops direct their surveillance efforts by ``following their nose''. As explained below, there is a stigma field in the neighborhoods which the cops use to choose where to patrol. We denote by $\theta$ the rate at which cops follow a stigmatized route.

\subsubsection*{The classifier.}
During this toy model we will keep the classifier as simple as possible. Indeed, we choose a random classifier with a jail sentencing rate of $r_c$. As unrealistic as this may seem, courts randomize assignment of defense lawyers and judges to defendants. In many cases the incarceration rates differ dramatically according to such assignment, allowing us to justify $r_c$ in this case as the probability of being assigned a good/bad lawyer and a lenient/strict judge \cite{anderson2012much}.

\subsubsection*{\bf Stigma Field.}
Finally, there is a variable at each point in space we call the ``Stigma Field''. It represents the bias towards a region where crime has been recorded. It is initialized to be zero.

\subsubsection*{\bf The Setup.}
The initial conditions can be described by the following parameters:

{\it Population size}. An original population size $N_G$ for each group. At the beginning we assume it to be the same across groups.

{\it Original configuration.}
At the beginning the agent population is randomly distributed across the grid points in their particular regions.

{\it Cop distribution.}
The cops are also placed at random but with a bias $q_0$: $8/10$ are placed in one region, while only $2/10$ are placed in the other. Note this is only their original configuration, they can still move between regions.

\subsubsection*{Dynamics}

The following are the simple rules agents follow during the arrest phase:

For population agents:
\begin{itemize}
\item Move to a neighboring cell at random.
\item With probability $c_0$, commit a crime.
\end{itemize}

For cops:
\begin{itemize}
\item With probability $\theta$, move to the neighboring cell with the highest value of the Stigma Field.
\item Face a random direction. With probability $\omega$, move $m_c$ steps, otherwise move one step.
\item Arrest agents in neighboring cells that have committed a crime this iteration.
\end{itemize}

Finally, the arrest process:
\begin{itemize}
\item Increase the Stigma Field at place of arrest by a given amount, at neighboring cells by smaller but nonzero amount.
\item Be judged by the classifier with hard $0/1$ assignment.
\item With probability $r_0$, recidivate.
\end{itemize}

Regarding the parameters of our particular model, we chose the crime rate to be $c_0=.01$. The recidivism rate was $r_0 = .4$; we chose recidivism rate this large not because it reflects truth but because it helped the model stabilize faster, smaller values also work. We also picked an initial population size of $100$ and a cop probability of moving away from its position $\omega = .1$ and $m_c = 3$.

\subsection*{A few derivations}\label{sec:extra_derivations}
Recall the classifier is random, with probability of positive outcome to be $r_c$. From this we notice
\begin{align*}
FPR_A(g) &= \mathbb{P}(J=1 | R=0, A=1, G=g) \\
& = \mathbb{P}(J = 1 | A=1, G=g) \\
& = r_c.
\end{align*}
Similarly 
\begin{align*}
FNR_A(g) & = \mathbb{P}(J=0 | R=1,A=1,G=g) \\
& = \mathbb{P}(J=0 | A = 1, G = g) \\
& = 1 - r_c,
\end{align*}
hence we achieve A-conditioned ERB. Second recall the agents recidivate with constant rate $r_0$, independent of classifier outcome, hence 
\begin{align*}
PPV_A(g) &= \mathbb{P}(R=1 | J=1, A=1, G=g) \\
& = \mathbb{P}(R=1| A=1,G=g) \\
&=r_0.
\end{align*}
Hence we achieve A-conditioned PP.

To show P-PPV and A-PPV are the same just note:
\begin{align*}
PPV_P(g) &= \mathbb{P}(R=1 | J=1, G=g) \\
& = \mathbb{P}(R = 1 | J=1, G=g, A=1) \mathbb{P}(A=1 | J=1,G=g) \\
&+ \mathbb{P}(R=1 |J=1,G=g,A=0)\mathbb{P}(A=0 |J=1,G=g) \\
& = \mathbb{P}(R = 1 | J=1, G=g, A=1) \times 1,
\end{align*}
since recidivism is not possible without an original arrest, and since only those arrested can be judged by the classifier.

\bibliographystyle{ACM-Reference-Format}

\end{document}